\documentclass{article}

\makeatletter
    \let\@fnsymbol\@alph
\makeatother
\usepackage{caption}
\usepackage{arxiv}

\usepackage[utf8]{inputenc} 
\usepackage[T1]{fontenc}    
\usepackage{hyperref}       
\usepackage{url}            
\usepackage{booktabs}       
\usepackage{amsfonts}       
\usepackage{nicefrac}       
\usepackage{microtype}      
\usepackage{lipsum}		
\usepackage{graphicx}
\usepackage{natbib}
\usepackage{doi}
\usepackage{multirow}
\usepackage{colortbl}

\title{A hybrid-model approach for reducing the performance gap in building energy forecasting}

\author{
    \href{https://orcid.org/0000-0001-8504-2303}{\hspace{1mm}Xia Chen $^*$ \thanks{Technische Universität Berlin, Germany,  Leibniz University Hannover, Germany <xia.chen@iek.uni-hannover.de>}} \\
	\And
	\href{https://orcid.org/0000-0002-8118-7422}{\hspace{1mm}Tong Guo\thanks{Hermann-Rietschel-Institut, Technische Universität Berlin, Berlin Germany}} \\
	\And
	\href{}{\hspace{1mm}Martin Kriegel\footnotemark[2]} \\
	\And
	\href{https://orcid.org/0000-0002-0935-4361}{\hspace{1mm}Philipp Geyer\thanks{Technische Universität Berlin, Germany, Leibniz University Hannover, Germany}} \\
}

\renewcommand{\shorttitle}{A hybrid-model approach for reducing the performance gap in building energy forecasting}

\hypersetup{
pdftitle={A hybrid-model approach for reducing the performance gap in building energy forecasting},
pdfsubject={},
pdfauthor={X Chen},
pdfkeywords={},
}

\begin{document}
\maketitle

\begin{abstract}
	The performance gap between predicted and actual energy consumption in the building domain remains an unsolved problem in practice. The gap exists differently in both current mainstream methods: the first-principles model and the machine learning (ML) model. Inspired by the concept of time-series decomposition to identify different uncertainties, we proposed a hybrid-model approach by combining both methods to minimize this gap: 1. Use the first-principles method as an encoding tool to convert the building static features and predictable patterns in time-series simulation results; 2. The ML method combines the results as extra inputs with historical records simultaneously, trains the model to capture the implicit performance difference, and aligns to calibrate the output. To extend this approach in practice, a new concept in the modeling process: Level-of-Information (LOI), is introduced to leverage the balance between the investment of simulation modeling detail and the accuracy boost. The approach is tested over a three-year period, with hourly measured energy load from an operating commercial building in Shanghai. The result presents a dominant accuracy enhancement: The hybrid-model shows higher accuracy in prediction with better interpretability; More important, it releases the practitioners from modeling workload and computational resources in refining simulation. In summary, the approach provides a nexus for integrating domain knowledge via building simulation with data-driven methods. This mindset applies to solving general engineering problems and leads to improved prediction accuracy. The result and source data is available at \href{https://github.com/ResearchGroup-G/PerformanceGap-Hybrid-Approach}{https://github.com/ResearchGroup-G/PerformanceGap-Hybrid-Approach}.
	
	\begin{table}[h]
    \centering
    \begin{tabular}{lll}
        \toprule
        ANN & Artificial Neural Network &  \\
        BIM & Building Information Modeling &  \\
        BPS & Building Performance Simulation &  \\
        COP & Coefficient of Performance &  \\
        HVAC & Heating, Ventilation, and Air Conditioning &  \\
        LightGBM & Light Gradient Boosting Machine &  \\
        LOD & Level-of-Development &  \\
        LOI & Level-of-Information &  \\
        ML & Machine Learning &  \\
        NRMSE & Normalized Root Mean Squared Error &  \\
        R² & R-squared &  \\
        RMSE & Root Mean Squared Error &  \\
        STL & Seasonal and Trend decomposition using Loess &  \\
        WWR & Window-Wall-Ratio & \\
        \bottomrule
    \end{tabular}
    \end{table}

\end{abstract}

\keywords{Building performance simulation \and Uncertainty \and Performance gap \and Hybrid-model
approach \and Level of information}

\section{Introduction}
With the global digitalization trend, two major changes have occurred in the building sector: (i) the boom of available data volume, especially in operation monitoring (dynamic time-series data) and building characteristics (static building features), (ii) the increasing reliance on building performance simulation (BPS) \citep{Hensen.2011} or predictive models to support energy-efficient design 
\citep{Rezaee.2019}. In this context, a performance gap between measured data and predicted output in two major modeling approaches: first-principles methods or white-box approaches (simulation tools) and data-driven models or black-box approaches (ML models), is reported \citep{Wilde.2014}.

The first-principles model reproduces the physical energy processes of buildings by equations \citep{Clarke.2007}. Numerous tools have been developed to simulate physical and thermal behaviors in buildings, e.g., TRNSYS \citep{klein2007trnsys}, EnergyPlus \citep{crawley2000energy}, etc. However, precise modeling and accurate results usually require detailed building characteristics with significant modeling effort, making it less cost-efficient in practice with full-scale experiments. Sequentially, without real-data support, it is difficult for BPS tools to simulate dynamic non-physical patterns or involve implicit factors in the modeling process, e.g., weather, user behavior, and occupancy \citep{Sonta.2018}. Furthermore, BPS tools normally rely on the parametric modeling process \citep{Sousa.2012}. Compared to building static characteristics (u-value, internal mass, etc.), historical records are rarely integrated for calibration or optimization \citep{Nguyen.2014}. 

The difficulties associated with the first-principles modeling process have contributed to developing alternative approaches based on data-driven methods. Especially, machine learning (ML) methods that enable designers to adopt energy models without being explicitly programmed have become popular in the recent decade \citep{Seyedzadeh.2018,Westermann.2019}. ML approaches perform a promising ability for capturing data hidden patterns by minimizing the defined mathematical loss function through the training process; They have been proved to conduct efficient and accurate results where historical records are available \citep{Chakraborty.2019, Deb.2017}. However, these black-box models are created directly from data by the pure mathematical process without considering the underlying physics of building thermal and energy systems, making ML models train in a relatively inefficient way by requiring a large amount of data for pattern learning. The current research on data-driven methods in building performance prediction focuses almost exclusively on pure ML methods, developing new ML approaches in feature engineering, model structure, and objective functions for accuracy enhancement \citep{Chakraborty.2019, Banihashemi.2017, Amasyali.2018}.

Furthermore, difficulties with accurate BPS are also associated with comprehensive building data collection. In general, more input parameters describe a better representation of building energy performance in detail, leading to a higher simulation accuracy \citep{Coakley.2011}. However, in some cases, huge impacts on the result are only caused by small underlying data sets \citep{Hsu.2015}, which makes the effort of collecting more data not closely correlated to the higher performance. The mainstream to collect and layer building information in the Building Information Modeling (BIM) is described by the Level-of-Development (LOD) \citep{Borrmann.2018}. Although this concept is widely accepted for building design \citep{Farzaneh.2019, Abualdenien.2018} and construction \citep{Latiffi.2015}, LOD is not fully applicable for the BPS. Although the influence of uncertain static feature information for BPS has been examined by \citep{Singh.2020}, combined evaluation with dynamic information from historical records is not addressed yet \citep{Gao.2019}. In this context, a proper approach to tackle the assessment of static and dynamic features within the scope of information-accuracy balance is missing.

Given this context, we recognized the current limitation of both mainstream methodologies and the great potential of combining them in our domain. The research questions that this paper aims to address are:
\begin{itemize}
\item \textit{Whether bridging first-principles approaches and data-driven methods in our domain can further reduce the performance gap. 
\item To implement the combined methods into the practice, how to evaluate and achieve the detail-accuracy balance?}
\end{itemize}

From the discussion above, both the ML (data-based) and the first-principles model (knowledge-based) are driven by different theoretical fundamentals with extensive support within the BPS community, yet rare research attempts for integration. To our best knowledge, only rare successful integration cases are reported: in the urban energy modeling scale \citep{Nutkiewicz.2018}, and for building occupant activities understanding \citep{Sonta.2018}. Inspired by the time-series decomposition, we firstly identified the different uncertainties in the process. Then, the presented approach develops a hybrid-model framework for BPS and prediction, which enable to capture of both systematic knowledge (by first-principles modeling) and implicit patterns (by ML induction) to improve the prediction accuracy. Furthermore, to leverage the balance between the investment of simulation modeling detail and the accuracy boost, a new concept specification: Level-of-Information (LOI), is proposed: LOI is used in evaluating contributions and balancing between modeling detail and ML accuracy boost. In this paper, we chose an operating commercial building from Shanghai as a case study to validate the method. 

The specific contributions that we conducted in this paper are:
\begin{enumerate}
\item We categorized uncertainties in the building performance gap between predictions and historical records. Furthermore, our research indicated that knowledge-based approaches and data-driven methods have complementary roles in minimizing different uncertainties. 
\item We proposed a method to allow knowledge-based approaches and data-driven models to integrate into a hybrid framework. The framework utilizes information efficiently from both static characteristics and dynamic historical records, which achieves a considerable accuracy improvement with less precise building details required. We applied this framework in the BPS domain to bridge the performance gap.
\item The new concept: Level-of-Information (LOI), is proposed to analyze the balance between information detail investment and accuracy boost for applying the hybrid-model approach in practice.
\end{enumerate}

To develop the hybrid approach using simulation and ML model, Section \ref{Methodology} introduces the theoretical support in time-series decomposition and uncertainties identification, methodologies of the first-principles model and the ML method used in the hybrid-model approach. Section \ref{Case study} describes the setup of a case study for validating the approach by data from a commercial building in Shanghai, China; Section \ref{Results} discusses the case study results. Section \ref{Discussion} outlines the limitations and future work, and Section \ref{Conclusion} concludes the paper.

\section{Methodology}
\label{Methodology}

\subsection{Time-series decomposition and uncertainty: systematic and unsystematic patterns}
\label{Time-series decomposition and uncertainty: systematic and unsystematic patterns}

\begin{figure}[h]
	\centering
	\includegraphics[width=16cm]{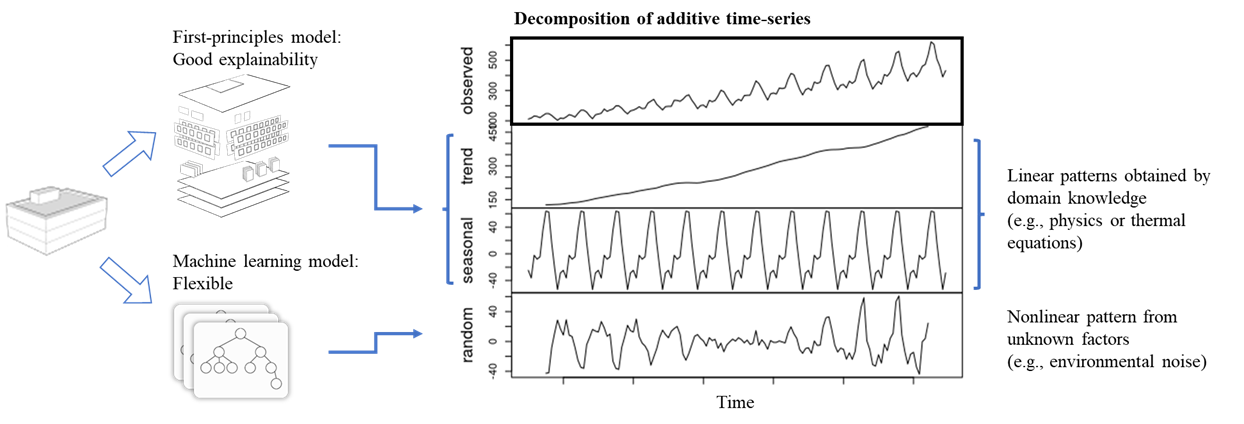}
	\caption{Analogous of STL decomposition; The building energy performance curve is equivalent to an additive time-series set of trend, periodicity (systematic), and residual (unsystematic) sub-series, which are better captured by first-principles models and ML models, individually. }
	\label{fig:Fig1.png}
\end{figure}

For building energy consumption, historical records compress all the information regarding building features, user behavior, operation conditions, as well as unpredicted factors into a compact time-series format. This compressed information-encoded format causes difficulty for the ML model to extract a variety of hidden patterns. In building energy assessment, we found one review paper to conduct different sources (parameters), methods, and processes to locate uncertainties \citep{Tian.2018}. However, we argue that it is necessary to distinguish them first by their innate characters, and address them accordingly afterward. In time-series forecasting, it is often helpful to split a time-series into several sub-series, each representing an underlying pattern category \citep{Hyndman.2018}. An informative and the most common decomposition method is Seasonal and Trend decomposition using Loess (STL) \citep{Cleveland.1990} by a purely mathematical process. It defines a series $y$ as an additive or multiplicative combination of trend ($T$), seasonal or periodicity ($P$), and residual or random ($R$) over time $t$. Figure \ref{fig:Fig1.png} visualizes this concept in the middle, in which the original time-series is illustrated in the first row as observed, accompanied by decomposed series underneath. 

In general, the STL decomposition includes an essential abstraction mindset to distinguish time-series data into systematic and unsystematic patterns: 
\begin{itemize}
\item Trend and periodic sub-series are related to systematic physical patterns, which have consistency and recurrence. They are suitable for knowledge-based simulation.  
\item The random sub-series is referred to as the stochastic part, which due to implicit factors or lack of information unable to be directly modeled. 
\end{itemize}

This mindset also implies the theory of \textit{uncertainty decomposition}, which refers to two distinct types of uncertainties: \textit{aleatory} and \textit{epistemic} \citep{DerKiureghian.2009}. Aleatoric uncertainties represent the inherent randomness from the natural data generation process and hence are less likely reduceable. Epistemic uncertainties are risen due to limited data or understanding, which is possible to be reduced by gathering more information or by refining models. Epistemic uncertainties are further decomposable into two sub-types: \textit{parametric} and \textit{structural} \citep{JeremiahLiu.2019}. In this context, we need to firstly identify epistemic uncertainties within the BPS, reduce them individually by assigning sub-types into first-principles approaches and ML models:

\begin{itemize}
\item The first-principles model is qualified to conduct deterministic processes based on the building decomposition knowledge and explained by physical laws of thermodynamics (trend and seasonal patterns). In addition, it offers domain insights to reduce epistemic uncertainties due to the ML modeling limit on structural specification (monolithic model without integration of domain knowledge explained by physics or thermal equations).
\item ML models are more efficient to induct from data and capture the stochastic hidden patterns via the mathematical training process. This process conducts information from historical records, including implicit factors, to supplement the lack of parametric specification from simulation’s limitation (e.g., ignorance of certain building design factors).
\end{itemize}

\begin{figure}[h]
	\centering
	\includegraphics[width=15cm]{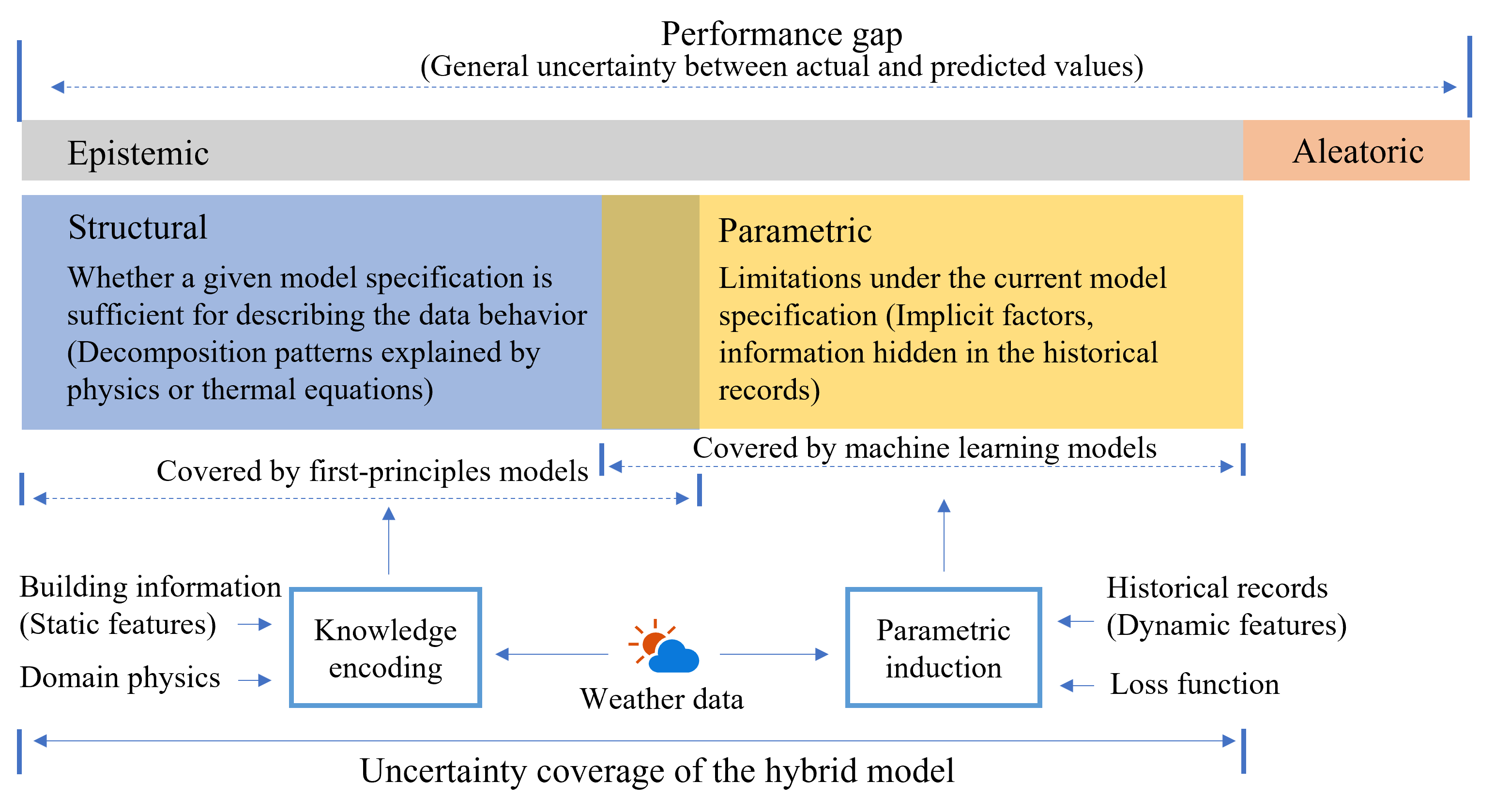}
	\caption{Uncertainty decomposition and the coverage range for different methods. The performance gap between actual records and predictions exists due to epistemic and aleatoric uncertainties in modeling. Within epistemic uncertainties, both first-principles models and ML models are complementary advantageous at covering structural and parametric uncertainties, individually; By combining both methods' strengths, i.e., the hybrid approach will have a border uncertainty coverage and further reduce the performance gap. }
	\label{fig:Fig2.png}
\end{figure}

The entire available information representation of a building, including static and dynamic parameters, is properly allocated into two approaches, integrated under the thought of uncertainty decomposition. Given these points, a summary illustration is presented in Figure \ref{fig:Fig2.png}. A process to combine these two approaches (first-principles models and ML models) is the critical key solution offered in our hybrid framework.

\subsection{First-principles model}
\label{First-principles model}

The first-principles model (or white-box models) in the building simulation domain requires modeling of thermodynamical processes, which relies on equations describing the physical behavior of thermodynamics and heat transfer. This paper chose IDA Indoor Climate and Energy (IDA ICE) as our first-principle model, a well-known dynamic simulation software to study energy consumption, indoor air quality, and thermal comfort in the building \citep{Milic.2018}. In this study, IDA ICE version 4.8 is implemented.

As mentioned in the introduction section, first-principles models require a complete description of the building to set up the simulation appropriately. To start with the modeling process, we need to consider the type of input parameters. In BPS, they are typically classified into five types: 
\begin{itemize}
\item Exterior conditions (e.g., weather data, shading by surrounding objects, urban details, etc.)
\item Basic geometry (e.g., floor area, number of floors, zoning information, etc.)
\item Buildings physics (physical features of building components: wall, constructions, materials, etc.)
\item Energy systems (e.g., HVAC, renewables, etc.)
\item Internal details (e.g., use conditions and user behavior, etc.)
\end{itemize}

With the underlying physical behavior equations, all these parameters with modeling knowledge incorporation provide insights into short-term patterns (day/night change; user behavior: workday/weekend change) and mid-/long-term trends (seasonal change; climate change), as \textit{periodicity} (\textit{P}) and \textit{trend} (\textit{T}). However, the difficulties in corresponding feature collection are the major challenge in the first-principles modeling process and the accuracy normally aligns with the increasing time consumption and resource investment. To further investigate this trade-off, we proposed a new concept specification: \textbf{Levels-of-Information (LOI)}. 

\subsection{Level-of-Information (LOI)}
\label{Level-of-Information (LOI)}

To intuitively represent the difference in available detail information, the refinement level of building geometric and semantic information is defined as Levels-of-Information (LOI). LOI is a concept specifically designed for hybrid model data layering: how different static building characteristics for the first-principles model contribute to the performance enhancement of ML models trained by historical records. 
Differ from LOD in BIM, the objective of LOI emphasizes evaluating required details and importance for the BPS, including all related variables, many of which may be unnecessary or redundant \citep{Hsu.2015}. The LOI categorization guideline refers to the three general data perspectives: expediency, transparency, and explainability \citep{Judea.2021}. To systematically assign variables in different LOIs, we proposed three core aspects for LOI evaluation derived from the commonality in domain literature: \textit{Implementation cost}, \textit{Importance level}, and \textit{Universality}.

\subsubsection{Implementation cost}

Implementation cost reflects the difficulty of obtaining the required data and its reliability.

\begin{figure}[h]
	\centering
	\includegraphics[width=12cm]{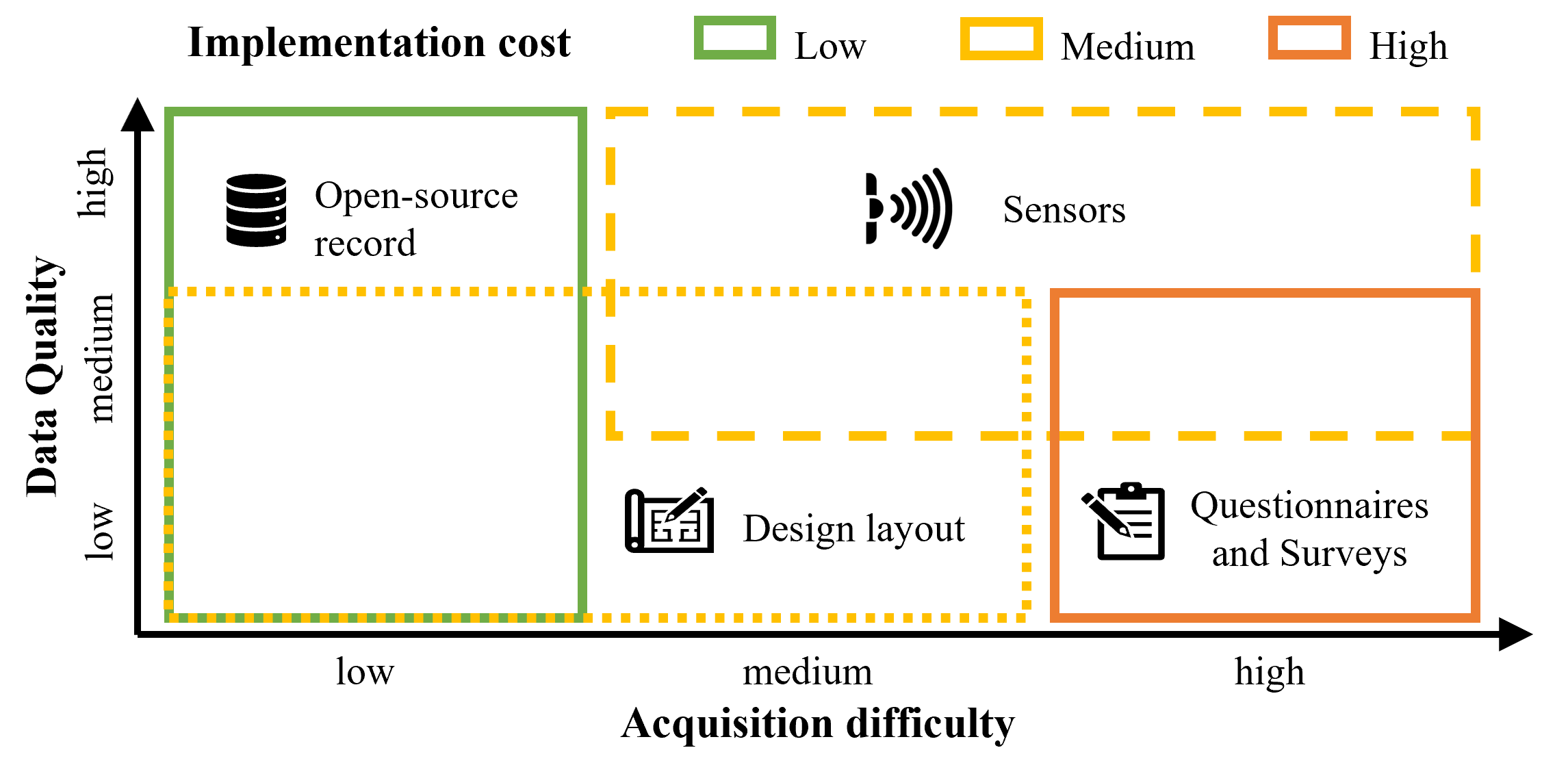}
	\caption{Representation of implementation cost for different data collection methods with respective data qualities; Some methods need to consider the implementation case, e.g., design layout, as well as sensors, have relatively lower data implementation costs for a single building than building stocks. }
	\label{fig:Fig3.png}
\end{figure}

Despite the exponential growth of building data volume in recent years, researchers find it difficult to directly utilize data from the real world owing to variances in data format, source, spatial resolution, and quality \citep{Monteiro.2018, Tian.2018}. To investigate the data implementation cost, we need to consider two aspects: \textit{the cost of gathering data} and\textit{ the effort for gaining good data quality}. Both aspects require a significant amount of time and labor. By implementing different parameters or processing methods, the data contains uncertainties regarding quality \citep{Tian.2018} at different levels. In this paper, we generally discuss four literature-derived primary sources for obtaining data from building: \textit{open-source records}, \textit{energy consumption data obtained by sensors},\textit{ structure information from construction plans}, and \textit{occupancy data through questionnaires} \citep{Mantha.2016}. A conceptual implementation cost illustration of different data acquisition difficulties and the data qualities is presented in Figure \ref{fig:Fig3.png}.

Compared to open-source, data from sensors contains more valuable information for modeling but is more challenging in acquisition. Other primary building information, including building materials, zones, and systems, are accessible from design layouts yet often not available in public. They need to be gathered from relevant design offices or construction units, leading to uneven data qualities and relatively serious acquisition difficulty. Even though technological advancement has enhanced behavior information gathering, data on occupant comfort and occupant attributes are only accessible via questionnaires. Besides, the information from questionnaires and surveys is often relatively subjective and unstable in quality. Depending on the data acquisition source and design cases, the same building feature may have different implementation difficulties.

\subsubsection{Importance level}

The importance level measures the necessity and sensitivity of certain input variables for the modeling during the simulation process. 

\begin{figure}[h]
	\centering
	\includegraphics[width=14cm]{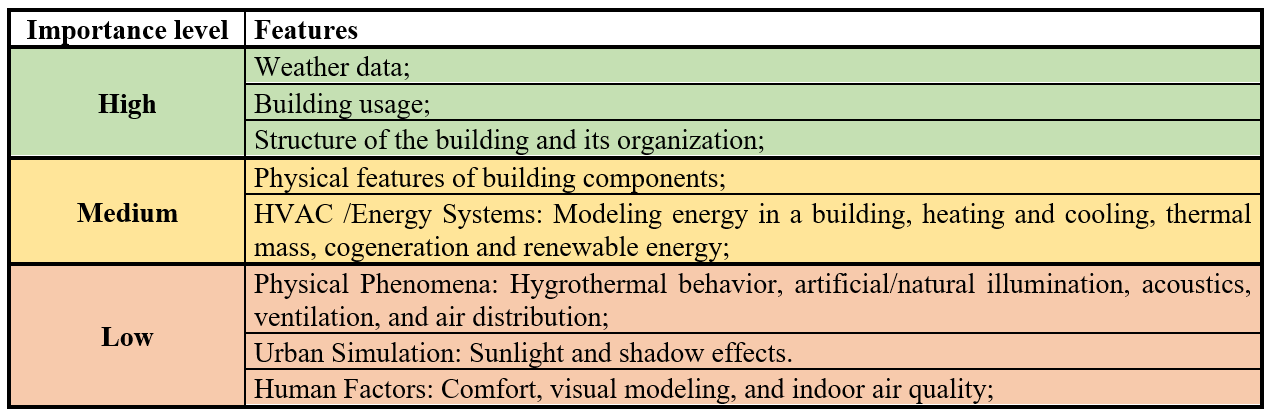}
	\caption{Importance level summary of features data.}
	\label{fig:Table1.png}
\end{figure}

Up to today, many pieces of research have been carried out on this topic and established a minimal number of critical feature sets for creating building energy models \citep{Jaeger.2020,Bilal.2016}. Recent research also examined the information required with sensitivity analysis for the building energy performance \citep{Singh.2020}. Based on the thorough reviews and corresponding literatures regarding building first-principles models \citep{Hensen.2011,stergard.2016,Wang.2017} and simulation tools \citep{Nguyen.2014,Kamel.2019}, we summarized the top three most important categories: structure of the building and its organization (geometry features), building usage, and weather data, following by two influential features: energy systems and physical features of building components. In addition, we listed out additional aspects depending upon the tools from the above-mentioned reviews, as shown in Figure \ref{fig:Table1.png}. 

\subsubsection{Universality}

Universality is involved to reflect the degree of usage frequency of different variables in the BPS community. Variables that have higher universality correspond to better acceptance, more methodologies or validation in the community, and better data availability. 

\begin{figure}[h]
	\centering
	\includegraphics[width=16cm]{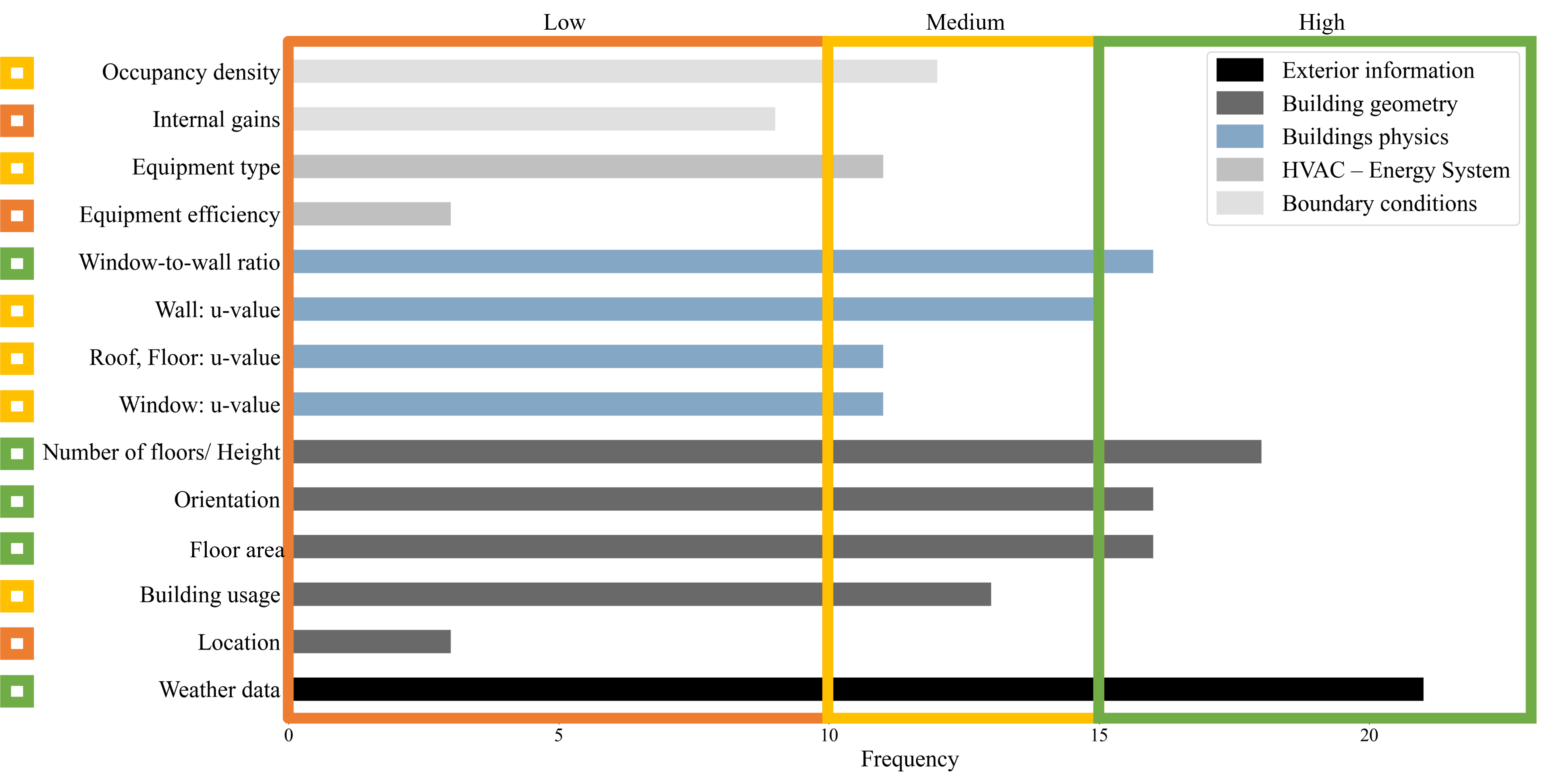}
	\caption{Statistic summary of universality based on review data \citep{Roman.2020}: Features are divided into three universality levels based on the frequency of occurrence from the review (Multiple occurrences of the same feature type in a paper are counted only once): Above 25 is marked as high (green). The value between 10 and 15 is the second level – medium (yellow), and the frequency less than 10 is marked as low (red); Features occurring less than 3 are not listed in the figure and are marked as low universality. }
	\label{fig:Fig4.png}
\end{figure}

A thorough statistic review for input variables selection has been done \citep{Roman.2020,Roman.2020b}. For reference, an overview of the frequency of building variables attribute in this review is recategorized and represented by the bar graph shown in Figure \ref{fig:Fig4.png}.

The inputs are classified into five main categories considering the \textit{exterior information} (climatic conditions), the \textit{building geometry} (interior dimension, zoning information, etc.), the \textit{buildings physics} (physical features of building components: walls, constructions, etc.), the \textit{energy system} (cooling, heating, ventilation, and air conditioning), and \textit{boundary conditions} (e.g., zone conditions and user behavior). These variable categories are represented by each row of the bar graph in Figure \ref{fig:Fig4.png}.  

\begin{figure}[h]
	\centering
	\includegraphics[width=16cm]{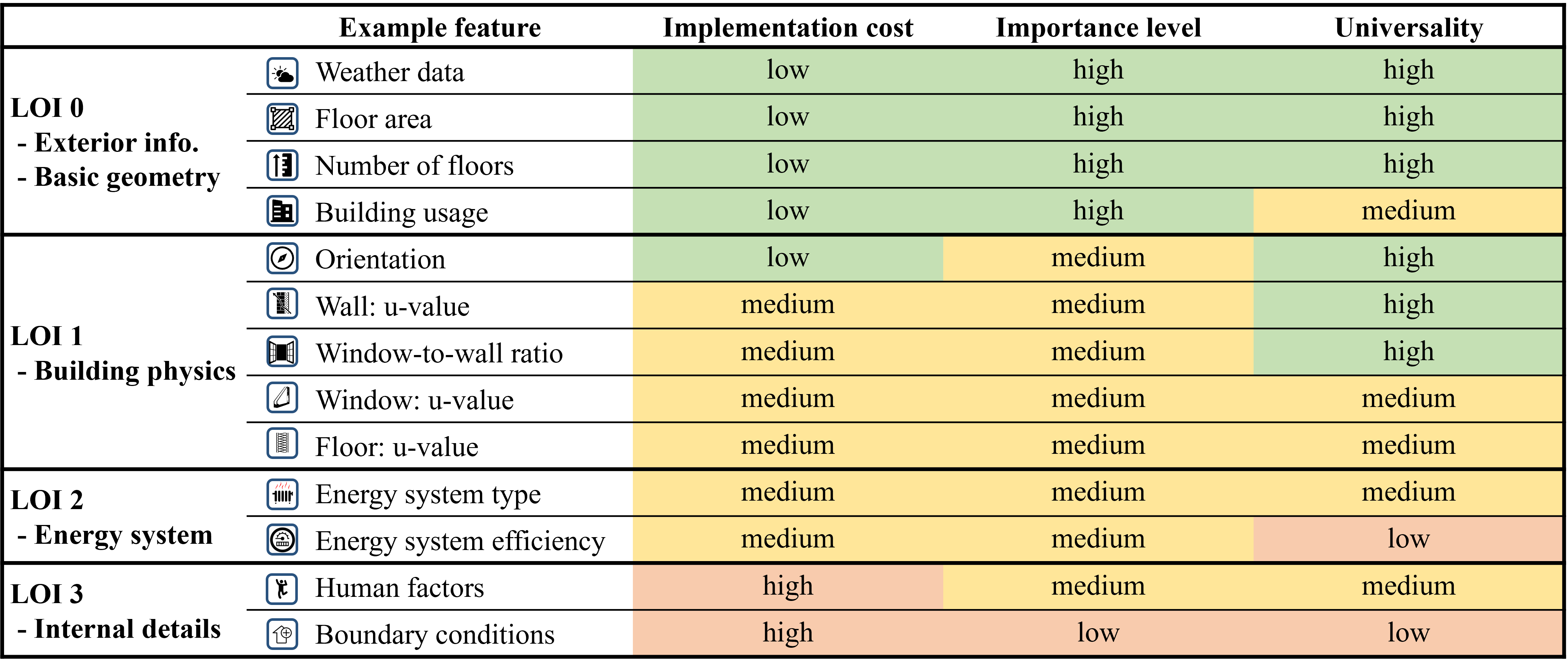}
	\caption{Four levels of LOI and example feature assignment based on three key evaluation aspects. }
	\label{fig:Fig5.png}
\end{figure}

By taking all the principles mentioned above into account, four LOIs are defined, as shown in Figure \ref{fig:Fig5.png}. We then allocated available building features into different LOIs: The default LOI-0 includes the most critical information for modeling, e.g., floor area, building height, building usage (operation schedule), and weather data, which are required at minimum to build the baseline physical model. The LOI-1 model involves more information regarding building physics, e.g., WWR (wind-wall-ratio) and other component details. The LOI-2 model further covers energy system information, including system type information and equipment efficiency details. Additional information on boundary conditions and human factors belongs to the final LOI-3. As the LOI level increases, the overall importance of the input parameters shows a downward trend, while the difficulty of obtaining continues to increase and the decreased universality. 

\subsection{Machine learning model: Ensemble tree-based model}
\label{Machine learning model: Ensemble tree-based model}

In general, the selected ML algorithm in the hybrid-model approach should fulfill two primary objectives: 1. The learning process should involve a decent mix sampling of inputs. 2. The mechanism of the target ML should focus on capturing the residual (nonlinear data). In this context, tree-based ensemble algorithms are fit for our targets and have been well accepted in the BPS community \citep{Papadopoulos.2018}. The idea behind this type of algorithm is to learn sequentially: The current regression tree is fitted to the residuals (errors) from the previous trees via boosting approach \citep{Marsland.2015}, which provides a better blending of different categories of data for learning. 

\begin{figure}[h]
	\centering
	\includegraphics[width=14cm]{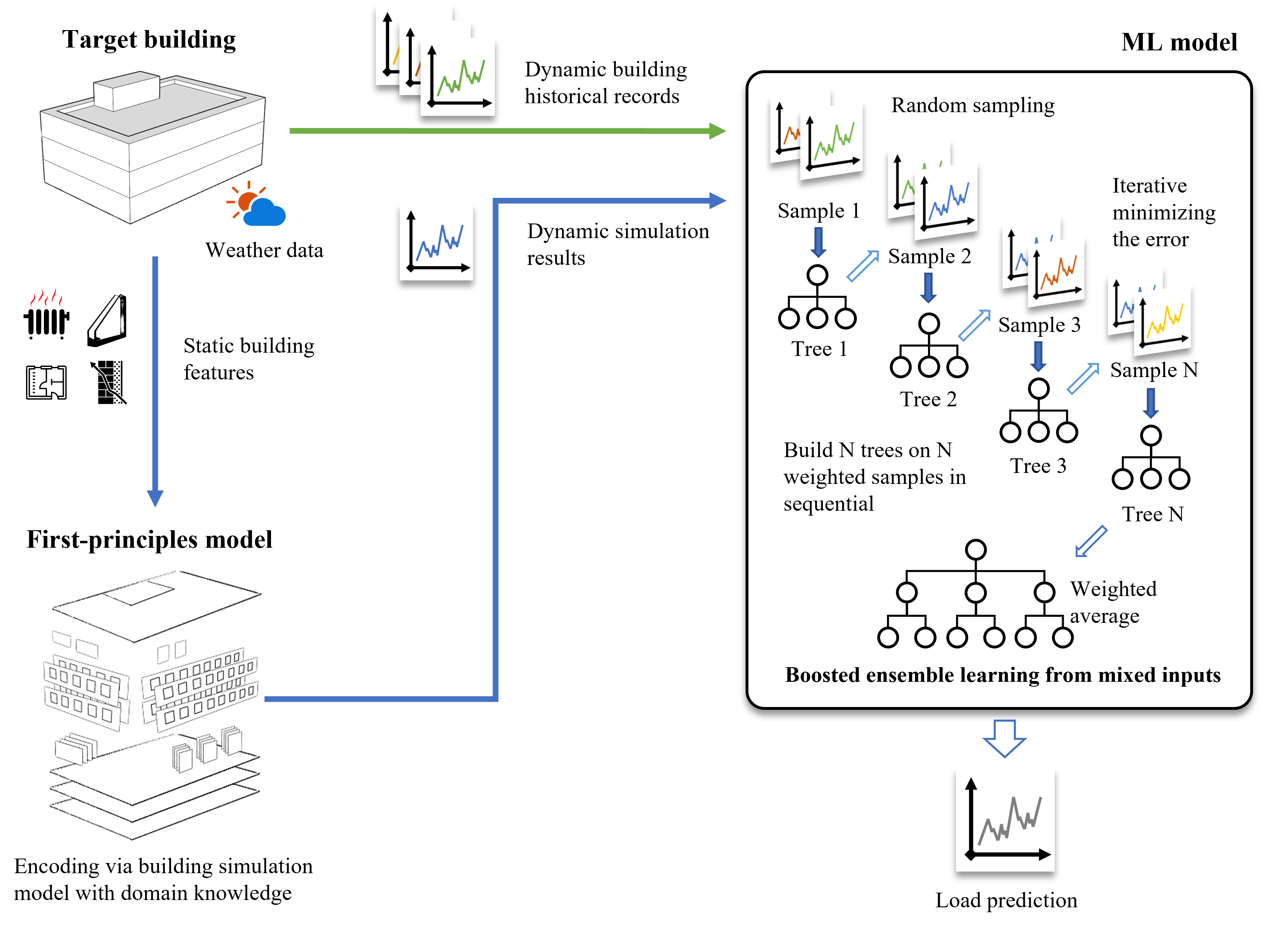}
	\caption{Demonstration of the hybrid approach process: the target building static features are firstly used in the first-principle model, combined with domain knowledge, and encoded into the simulation result; The result is then aligned with the historical records and weather data, fed into the ensemble tree-based model, the right box illustrated the process of mixture input data with the ensemble learning detail. }
	\label{fig:Fig6.png}
\end{figure}

In the ML community, ensemble algorithms are outperformance with their promising result and effective advancement for handling forecasting tasks in recent \citep{Arjunan.2020,Polikar.2006}. In the BPS domain, a comparative analysis reported that the ensemble learning model produces more accurate results than ANN and ordinary least square regression in a synthetic database from EnergyPlus simulations \citep{Chakraborty.2019}. Additionally, because the split finding mechanism in the ensemble tree-based algorithm is insensitive to the value range, data scaling is not required in data preprocessing \citep{Marsland.2015}. In this paper, we chose a well-known ensemble algorithm - LightGBM \citep{GuolinKe.2017,Chen.201639} as our ML model. A python-based LightGBM open-sources package is available\footnote{https://github.com/microsoft/LightGBM}. The illustrative hybrid-model approach process with boosting mechanism demonstration is presented in Figure \ref{fig:Fig6.png}.  

As seen in Figure \ref{fig:Fig6.png}, the output conducted by the first-principles model carries systematic patterns: encoded domain knowledge regarding thermodynamic behaviors, building characteristics, and component decomposition information (\textit{T} and \textit{P}) in time-series format. This representation aligns with the format of historical records, which enables the knowledge mentioned earlier to feed into ML models for further integration. The ML model then learns the difference between historical records and simulation output - the unsystematic patterns (\textit{R}) or parametric uncertainties caused by implicit factors in reality. The overall hybrid approach contributes to accuracy enhancement and model interpretation.

\section{Case study}
\label{Case study}

We designed the hybrid approach and firstly observed the performance novelty from the participation in a BPS competition \citep{TongXiao.2022} - a series of real-world commercial building energy prediction cases study in Shanghai, China. All required inputs for the method are collected from reliable sources, e.g., construction plans, equipment brochures, historical records, and national weather datasets. In this paper, we selected the test set building from the competition as our case. The target building has comprehensive static information records of geometric and component characteristics, as well as HVAC equipment parameters. The building installed energy monitors and thermal energy sensors, which collected historical energy usage data over a three-year period hourly, as well as corresponding weather data of Shanghai. 

For comparison, we trained three types of models for predicting air-conditioning power consumption: \textit{pure ML approach} (based on historical records), \textit{simulations} (first-principles model) \textit{in different building LOIs}, and corresponding levels of \textit{hybrid-model approaches}. All models’ performances are tested in the same set of scenarios: an entire year average, as well as selected two typical and two extreme periods. The case and method implementation details are expanded in the following subsections.

\subsection{LOI categorization}
\label{LOI categorization}

The available static building features for our case study are represented and categorized based on the criteria of LOI from Figure \ref{fig:Fig5.png} in Table \ref{tab:table1}. We further assigned all features into 12 different sub-levels for first-principles simulations. Higher LOI results in better chances of high accurate performance while requiring more effort for the modeling at the same time.

\begin{table}
\caption{LOI categorization in the case study; each sub-level is composed of basic features and extended features; for example: LOI 1-1 uses features from LOI 0-2 (Location, Number of floors, Floor area; Building usage) with the Orientation information as extended features.}
\label{tab:table1}
\centering
\arrayrulecolor{black}
\begin{tabular}{!{\color{black}\vrule}l!{\color{black}\vrule}l!{\color{black}\vrule}l!{\color{black}\vrule}l!{\color{black}\vrule}} 
\hline
\textbf{Level}                           & \textbf{Sub-level} & \textbf{Basic features}           & \textbf{Extended features}                                                                                                                                                                        \\ 
\hline
\multirow{2}{*}{\textbf{LOI 0}\textit{}} & \textbf{LOI 0-1}   & None\textit{}                     & \begin{tabular}[c]{@{}l@{}}•~~ Location: \textit{Shanghai, China}; \\•~~ Number of floors: \textit{30 + 4 (underground);} \\•~~ Floor area: \textit{101,806 m²}\textsuperscript{ };\end{tabular}  \\ 
\cline{2-4}
                                         & \textbf{LOI 0-2}   & LOI 0-1\textit{}                  & •~~ Building usage: \textit{Office};                                                                                                                                                              \\ 
\hline
\multirow{4}{*}{\textbf{LOI 1}\textit{}} & \textbf{LOI 1-1}   & \multirow{4}{*}{LOI 0-2\textit{}} & •~~ Orientation: \textit{30° east};                                                                                                                                                               \\ 
\cline{2-2}\cline{4-4}
                                         & \textbf{LOI 1-2}   &                                   & \begin{tabular}[c]{@{}l@{}}•~~ WWR: \textit{0.47}; \\•~~ u-value, window: \textit{0.4~ W/(m²K)};\end{tabular}                                                                                     \\ 
\cline{2-2}\cline{4-4}
                                         & \textbf{LOI 1-3}   &                                   & \begin{tabular}[c]{@{}l@{}}•~~ u-value, roof: \textit{0.62 W/(m²K)};  \\•~~ u-value, external wall: \textit{0.94 W/(m²K)}; \\•~~ u-value, basement ground: \textit{0.86 W/(m²K)};\end{tabular}    \\ 
\cline{2-2}\cline{4-4}
                                         & \textbf{LOI 1-4}   &                                   & •~~ Orientation + WWR +
  u-values;                                                                                                                                                               \\ 
\hline
\multirow{3}{*}{\textbf{LOI 2}\textit{}} & \textbf{LOI 2-1}   & \multirow{3}{*}{LOI 1-4\textit{}} & •~~ HVAC type: \textit{Full air volume system};                                                                                                                                                   \\ 
\cline{2-2}\cline{4-4}
                                         & \textbf{LOI 2-2}   &                                   & •~~ Chillers’ COP: \textit{5.59};                                                                                                                                                                 \\ 
\cline{2-2}\cline{4-4}
                                         & \textbf{LOI 2-3}   &                                   & \begin{tabular}[c]{@{}l@{}}•~~ HVAC type + Chillers’ COP \\•~~ Operating time: \textit{7:00 – 20:00};\end{tabular}                                                                                \\ 
\hline
\multirow{3}{*}{\textbf{LOI 3}\textit{}} & \textbf{LOI 3-1}   & \multirow{3}{*}{LOI 2-3\textit{}} & •~~ Human activity: \textit{medium};                                                                                                                                                              \\ 
\cline{2-2}\cline{4-4}
                                         & \textbf{LOI 3-2}   &                                   & •~~ Number of occupancies: \textit{15-20 m²/Person};                                                                                                                                              \\ 
\cline{2-2}\cline{4-4}
                                         & \textbf{LOI 3-3}   &                                   & •~~ Human activity +
  Number of occupancies;                                                                                                                                                     \\
\hline
\end{tabular}
\arrayrulecolor{black}
\end{table}

\subsection{Data processing}
\label{Data processing}

As mentioned in Section \ref{Machine learning model: Ensemble tree-based model}, we chose LightGBM as our method for both hybrid-model and pure ML approaches. Data preprocessing is required for model training. In this context, four different time-series features are involved: weather, time, historical records, and simulation output. To strengthen the periodicity information, we processed the time features by:
\begin{itemize}
\item Decomposing time features such as the month, week, day, etc.
\item Using sine/cosine transform for hour features.
\item Using the Boolean value to represent whether the day is a weekend or workday. 
\end{itemize}
More importantly, we used feature engineering methods from time-series forecasting to capture serial correlation of dynamic time-series features for the ML approach by:
\begin{itemize}
\item Shifting features for 1, 2, and 3 periods
\item Rolling average features for 6 and 12 time-windows with shifting 1, 3, 6, 12 periods
\end{itemize}

\subsection{Model training}
\label{Model training}

We kept the default setting of most hyperparameters\footnote{Default hyperparameter setting is available on LightGBM 3.2.1.99 documentation (2021): https://lightgbm.readthedocs.io/en/latest/Parameters.html?highlight=default} in LightGBM models but fine-tuned the iteration round to prevent the model under- or over-fitting: A 3-fold cross-validation is used to determine the best iteration. Both hybrid approach and pure ML model are implemented the same feature engineering.

\begin{figure}[h]
	\centering
	\includegraphics[width=16cm]{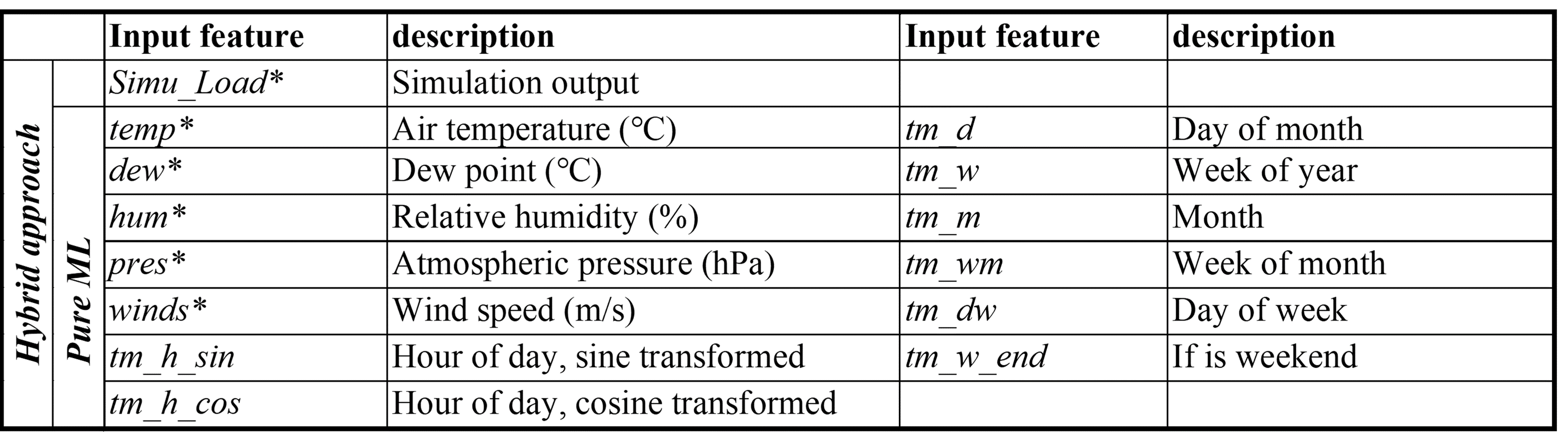}
	\caption{Input features for hybrid approaches and pure ML method (shifting and rolling features excluded).}
	\label{fig:Table3.png}
\end{figure}

The input difference between the pure ML model and the hybrid model lies wherever the simulation result (encoded knowledge) is involved as extra input features ($Simu_Load$ with feature engineering). The detailed information on input features is demonstrated in Figure \ref{fig:Table3.png}.

To prevent data leakage, we split the historical records into two parts: the first two years for training and the final years for validation. Subsequently, four different periods in the final year are selected for the study, as shown in Table \ref{tab:table2}. All approaches are fed-in or applied to the same data, including the pure ML method, first-principles models, and hybrid-model approaches in different LOIs.

\begin{table}
    \caption{Selected periods for hourly energy consumption prediction in our case.}
    \centering
    \begin{tabular}{llll}
        \toprule
        Interval & Days & Interval & Days \\
        \hline
        Winter typical & January 05 – January 17 & Summer typical & July 6 – July 18 \\
        Winter extreme & February 09 – February 21 & Summer extreme & July 22   – August 03 \\
        \bottomrule
    \end{tabular}
    \label{tab:table2}
\end{table}

\subsection{Metrics}
\label{Metrics}

Two common metrics are used for evaluating the accuracy of different LOI models: R-squared (R²) and Normalized Root Mean Squared Error (NRMSE). The coefficient of determination R² gives the proportion of the variance in measured data. Since R² is sensitive to extreme values but less sensitive to additive and proportional differences between simulated and measured data \citep{Vogt.2018}, NRMSE is added as a supplement to evaluate the accuracy of BPS models. NRMSE is the normalized version by the range of the dependent data variable and expressed in the percentage of RMSE, a quadratic scoring rule that measures the average magnitude of the error. It gives a relatively high weight to large errors. With the number of n observations in total, the \(y_i\) and \(\hat{y_i}\) mean the \(i^{th}\)  observed and predicted values, respectively. \(y_{min}\) and \(y_{max}\)  stand for the min and max of observed values. The formula of NRMSE is presented below: 

\begin{equation}
NRMSE =\frac{\sqrt{\frac{1}{n} {\textstyle \sum_{n}^{i=1}} (y_{i}-\hat{y_{i}})^{2} } }{y_{max}-y_{min}}*100\% 
\end{equation}

\section{Results}
\label{Results}

\begin{figure}[h]
	\centering
	\includegraphics[width=16cm]{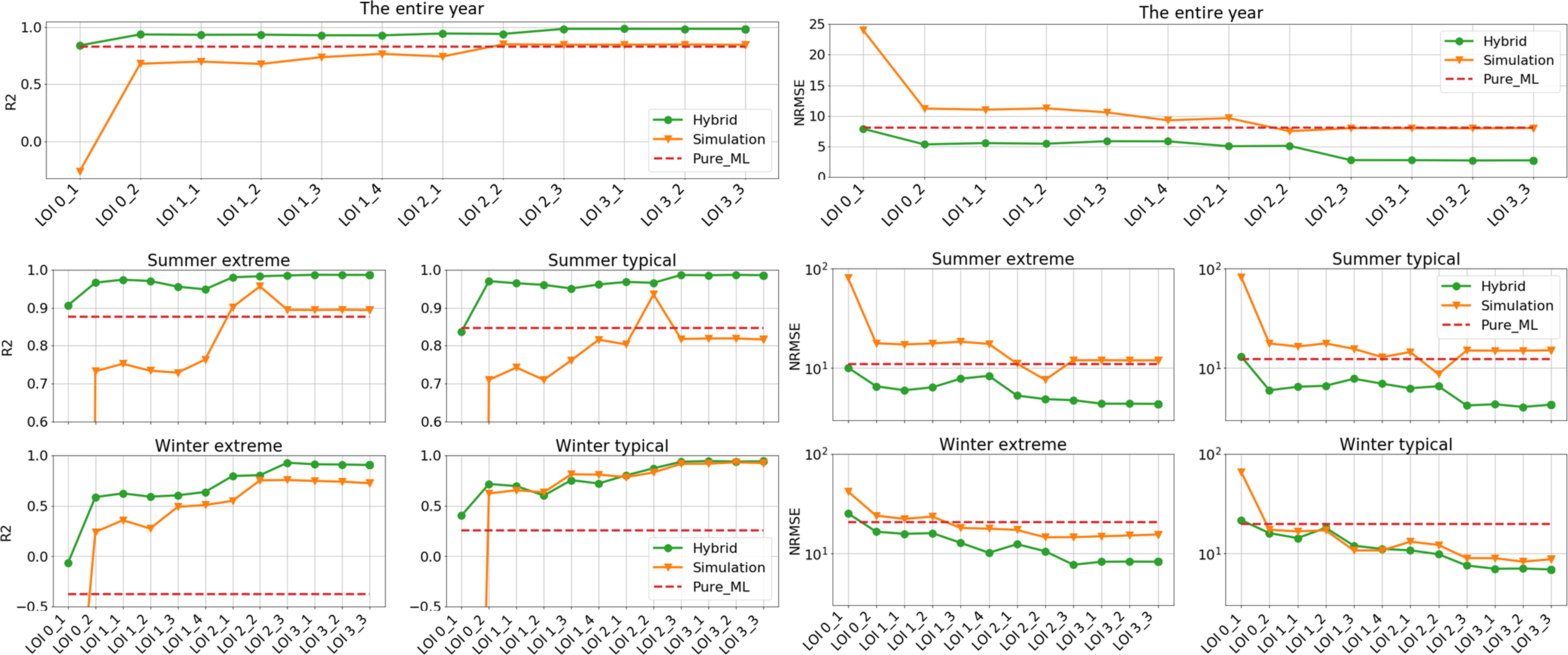}
	\caption{Comparison of results in different LOIs and periods under evaluation metrics of R-squared and NRMSE. }
	\label{fig:Fig7.png}
\end{figure}

Figure \ref{fig:Fig7.png} visualizes the performance of pure ML models, simulations in different LOIs, and corresponding hybrid-model approaches for the entire year and four selected periods under the evaluation metrics of R² and NRMSE, individually. As mentioned in Figure \ref{fig:Table3.png}, the pure ML model only contains time features and historical records for inputs. Therefore, the performance of pure ML models is not affected by different LOIs and remains constant.

From the entire year’s performance, hybrid-model approaches present a considerable accuracy boost (R² boosts from 0.75 to 0.97 on average from Figure \ref{fig:Fig7.png}) compared to simulations, as well as to the pure ML model in all LOI scenarios. This result validates two major points: 
\begin{enumerate}
\item The effectiveness of the hybrid approach under the concept of uncertainty decomposition from Figure \ref{fig:Fig2.png}. 
\item A proof of complementary advantage in reducing uncertainties between first-principles methods and ML models for building performance prediction.
\end{enumerate}

To sum up, the input from the first-principles method contributes to providing the ML model systematic information for further minimizing the forecasting error. Even the simulation at the LOI 0-1 (only basic geometric information fed in) with poor accuracy, the hybrid model still extracts valuable information from the simulation result. In all LOI 0-1 scenarios, hybrid-model approaches present a robust and higher accuracy than simulations and are better than the pure ML model in most selected periods. 

For the information-accuracy balance, the performance is generally enhanced along with higher LOIs. In the entire year’s performance, the simulation reaches the breakpoint and outperforms the pure ML at LOI 2-3 when basic geometric information, building type, building components’ features, the type of energy system, and the efficiency of the energy system are included. After the breakpoint, simulation accuracy in some cases experiences a decrease. With more details fed into simulations, the modeling reacts against inaccuracies of information added in LOI 3, the implicit factors, and the poor modeling approach in LOI 3 is sensitive. 

In summer typical, summer extreme, as well as winter extreme, there are significant lifts of accuracy present over most of the LOI in Figure 7. The overall R-squared of the hybrid model reaches nearly 0.98 at LOI 3-3 (full details), while the other two approaches are merely above around 0.8. The simulation performs with higher accuracy during the winter periods. Among different LOIs, the hybrid approach experiences striking accuracy enhancements when the information input reaches LOI 2-3 (LOI 2-2 plus operating time), which, in our opinion, is a balance point between the accuracy and the model detailing.
From the result of Figure \ref{fig:Fig7.png}, we conclude some noticeable points with interpretation: 

\begin{enumerate}
\item Observed from the performance in simulations and hybrid methods, the accuracy from LOI 0-1 (only basic geometric information) to LOI 0-2 has been dramatically improved. The corresponding added information at LOI 0-2 is the building usage setting, which contains the standard operating schedule in the simulation tools’ default template. By eliminating the uncertainty in building operation, the fixed operating schedule leads to a significant accuracy improvement.
\item An accuracy spike is observed during the summer period in LOI 2-2. The reason might be due to the default energy system setting mixed with the chiller’s COP, which causes a significant impact on the uncertainty of the building energy performance. By involving more information from Table \ref{tab:table1}, the spike vanished in the higher LOIs. Therefore, we believe that is rather a random, casual event in this case.
\item In terms of overall comparison, the accuracy with ML in summer is generally better. The reason lies in the building energy system: In the target building, cooling in summer uses the air conditioning system by consuming electricity only, while the heating supply in winter is covered via different facilities; hence, it contains more uncertainties in electricity consumption.
\end{enumerate}

\begin{figure}[h]
	\centering
	\includegraphics[width=15cm]{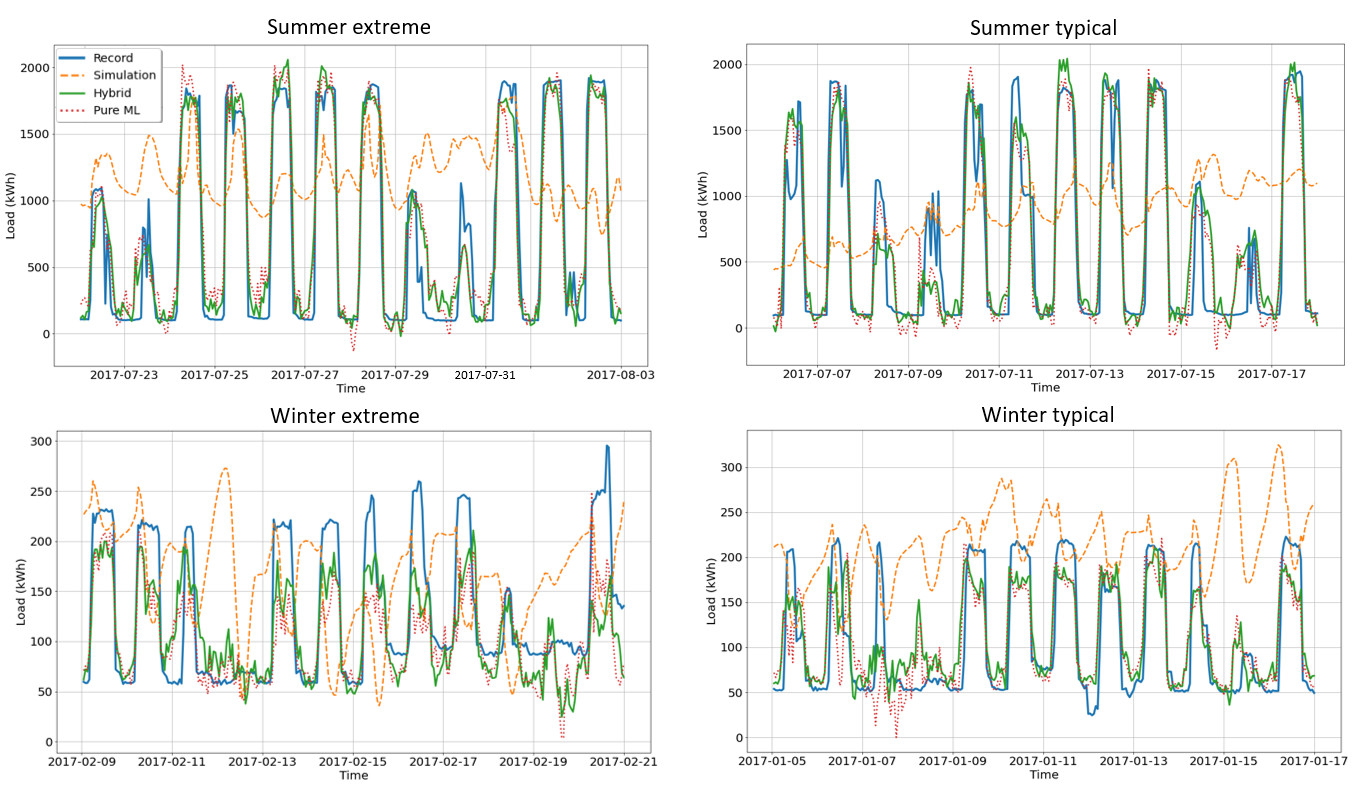}
	\caption{Performance comparison at four different periods – LOI 0-1; With only geometric information input without building usage specification, simulation results show relative imprecise in terms of numerical results. Pure ML has relatively high accuracy even with low LOI. However, there exists negative overshooting due to missing physical knowledge. This overshooting is further corrected by the hybrid-model approach. }
	\label{fig:Fig8.png}
\end{figure}

\begin{figure}[h]
	\centering
	\includegraphics[width=15cm]{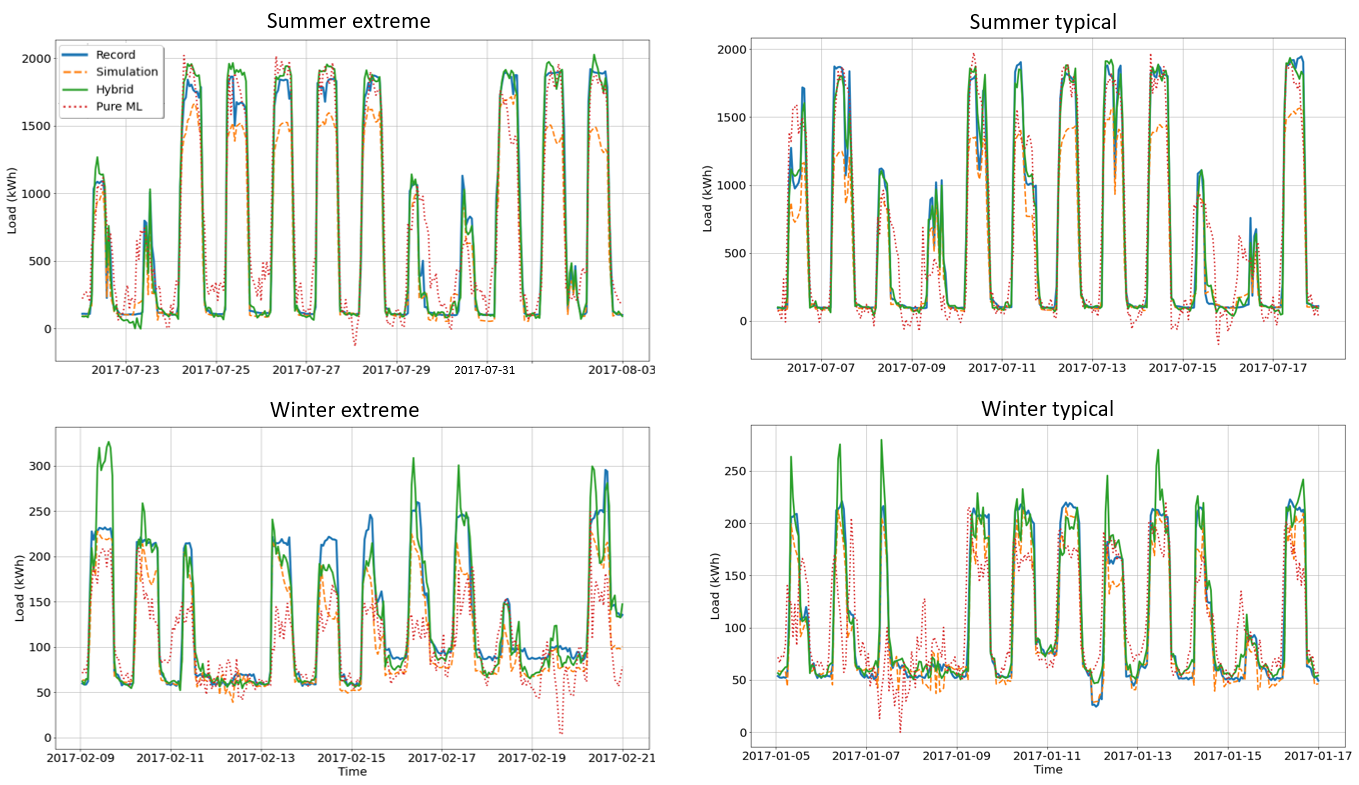}
	\caption{Performance comparison at four different periods – LOI 3-3; The simulation result contains the precise periodicity patterns that greatly smooth the fluctuations in baseload. The hybrid-model approach presents accuracy improvements by better covering peaks.  }
	\label{fig:Fig9.png}
\end{figure}

To further investigate the details of the accuracy improvement in the hybrid model, we selected the model from LOI 0-1 (least detail, only basic geometric information) and LOI 3-3 (full detail), visualized the prediction results through each of the four typical periods, as shown in Figure \ref{fig:Fig8.png} and Figure \ref{fig:Fig9.png}.

As presented in Figure \ref{fig:Fig8.png}, the performance of first-principles methods is rather inaccurate without the specification of the building type. The main difference between Figure \ref{fig:Fig8.png} and Figure \ref{fig:Fig9.png} is shown in the results of the first-principles methods and contributions to corresponding hybrid-model approaches. The simulation performances show that more detailed information modeling results in enhancements: better peak load, baseload, and periodicity patterns estimation accuracy with performance consistency. For example, from historical records, we observed regular pulses of five high peak loads with one or two low peaks, which stand for the working day and weekend load patterns. The first-principles method is impossible to capture the low peak loads by only relying on the law of physics and thermodynamics. By contrast, the hybrid-model approach obtains the flexibility of the ML model to capture the low peak loads more accurately (weekend loads usually contain more unknown factors, which is hard for parametric simulation). For LOI 1-1, the basic simulation result still contains day/night load patterns. This information via the hybrid-model approach integrated into ML methods is beneficial for performance enhancement: It offsets the major weakness of pure ML models - instability performance and negative overshooting at baseload; however, we realized that the implicit information from detailed simulation results does not always contribute to the hybrid approach performance positively: Compared to LOI 1-1, hybrid approaches at LOI 3-3 appear to overestimate the peak load, especially in the winter period.

\section{Discussion}
\label{Discussion}

In this section, the adaptability, limitations, and research potential regarding the framework, uncertainties categorization, and LOI are discussed. 

In the task of building energy consumption prediction, the hybrid-model approach performs with considerable accuracy boost while containing certain flexibility. We’d like to point out that the concept of this hybrid-model approach is universally applicable to general engineering simulation scenarios. Within the scope of the energy domain, this hybrid-model approach mindset with the advantage of accuracy and flexibility provides a solid cornerstone for developing future digitalization business models, e.g., dynamic demand-side control, better power market pricing signal, local district prosumer portfolio management, etc. For the limitation, the key is to find an adequate amount of effort for the model implementation, especially when the current widespread problem in the BPS community remains unsolved for data-driven methods: missing data and access difficulty in the real world. To increase attention and to expand acceptance of data-based methods in the industry, further effort should focus on broad performance validation, domain assistance system design, and database construction based on the hybrid approach, case by case. 

As for the uncertainty analysis in the scope of this paper, we only categorized and assigned it into different approaches for performance gap minimization. Uncertainties inherited in the building characteristic representations are worth further exploration. For example, if we try to specify uncertainties in the u-values or human activities, how to recognize and quantify the aleatory and epistemic uncertainties? In our case study, we observed a phenomenon that is worth mentioning yet hadn’t been raised attention and discussion in the community: the increase of information (which more commonly happens at high LOIs) could cause an accuracy decrease. Based on our research in the case data, we’ve concluded one of the possible explanations: the additional high LOI information with usually more serious implementation cost contains inaccurate information via assumption or estimation to fulfill the simulation data feed in. This misleading information results in an accuracy decline. In this context, whether given information will contribute to the model accuracy due to inherent uncertainty needs to be considered and further investigated. For an improvement, more cases studies, detailed refinement of the modeling process of first-principles models, and data governance are worth further research.

\section{Conclusion}
\label{Conclusion}

In this paper, we proposed a hybrid-model approach to demonstrate how domain knowledge via simulation incorporated with data-driven methods, especially ML leads to improved predictions. The case study has validated its novelty in the BPS domain: It accurately represents the cyclical variation while containing the flexibility to capture implicit patterns from historical records. The simulation output in time-series provides a proper format to embody static building features with domain knowledge into ML models. This idea of time-series-based decomposition further reduces the uncertainty gap between prediction and measured data. Compared to full-detailed physical simulation, the hybrid-model approach achieves higher accuracy with less precise building details required. Consequently, it is expected that the approach releases professional practitioners from modeling workload, time investment and computational resources, but only uses the first-principles model to extract sufficient systematic, physically explainable patterns of building energy behavior.

We further defined a new concept specification: Level-of-Information (LOI), to leverage the balance between the investment of simulation modeling detail and the accuracy boost. Based on the case study, the hybrid-model approach delivers a more reliable output compared to pure ML models or simulations. Several valuable pieces of information are summarized therein: 

\begin{enumerate}
\item For BPS, pure ML models without integrated domain knowledge are less efficient and flexible in extracting mixed information from historical records. Due to the absence of domain knowledge, the output of the pure ML contains noise that manifests itself on the baseload. Even a basic level of the first-principles approach has more promising performance than the pure ML model trained from dynamic historical records. 
\item Systematic information such as trend (T) and periodicity (P) are well predicted through domain knowledge modeling approaches. ML models present the ability to well capture implicit factors (R) by training to learn the difference between simulation output and historical records.
\item With additional building physics (LOI-1) and energy system (LOI-2) features from the simulation, the performance of the hybrid-model approach is greatly enhanced compared with basic geometry (LOI-0). From the case study conclusion, the trade-off between information detailing and accuracy contribution lies in LOI 2-3, including features in basic geometric information, building usage, building components’ features, the type of energy system, energy efficiency, and operating schedule. 
\item A promising path to integrate domain knowledge methods with data-driven models. Minimize the gap between predicted and measured performance guided by the concept of uncertainty decomposition.
\end{enumerate}

The performance gap and the difference between the first-principles method and the ML model (including deep learning) is a miniature of a long-existed debate: Which is better to account for human cognition and more promising, symbolism or connectionism? The hybrid approach opens new opportunities for future paradigms: knowledge-integrated machine learning. The symbolism, represented by the first-principles simulation in the case, allows for the integration of domain knowledge in ML with its underlying connectionism. This hybrid approach results in knowledge-based and data-driven prediction models that exploit the benefits of both sides, showing clear advantages compared to approaches relying only on one of the methods.

\subsection{Acknowledgements}
We gratefully acknowledge the German Research Foundation (DFG) support for funding the project under grant GE 1652/3-2 in the Researcher Unit FOR 2363. We would like to thank Prof. Peng Xu and his research group at Tongji University, Shanghai, China, for data resources support.

\bibliographystyle{unsrtnat}
\bibliography{references}  

\begin{thebibliography}{48}
\providecommand{\natexlab}[1]{#1}
\providecommand{\url}[1]{\texttt{#1}}
\expandafter\ifx\csname urlstyle\endcsname\relax
  \providecommand{\doi}[1]{doi: #1}\else
  \providecommand{\doi}{doi: \begingroup \urlstyle{rm}\Url}\fi

\bibitem[Hensen and Lamberts(2011)]{Hensen.2011}
Jan Hensen and Roberto Lamberts.
\newblock \emph{Building performance simulation for design and operation}.
\newblock {Spon Press}, Abingdon, Oxon and New York, NY, 2011.
\newblock ISBN 9781134026357.

\bibitem[Rezaee et~al.(2019)Rezaee, Brown, Haymaker, and
  Augenbroe]{Rezaee.2019}
Roya Rezaee, Jason Brown, John Haymaker, and Godfried Augenbroe.
\newblock A novel inverse data driven modelling approach to performance-based
  building design during early stages.
\newblock \emph{Advanced Engineering Informatics}, 41:\penalty0 100925, 2019.
\newblock ISSN 14740346.
\newblock \doi{10.1016/j.aei.2019.100925}.

\bibitem[de~Wilde(2014)]{Wilde.2014}
Pieter de~Wilde.
\newblock The gap between predicted and measured energy performance of
  buildings: A framework for investigation.
\newblock \emph{Automation in Construction}, 41:\penalty0 40--49, 2014.
\newblock ISSN 09265805.
\newblock \doi{10.1016/j.autcon.2014.02.009}.

\bibitem[Clarke(2007)]{Clarke.2007}
Clarke.
\newblock \emph{Energy Simulation in Building Design}.
\newblock Routledge, 2007.
\newblock ISBN 9780080505640.
\newblock \doi{10.4324/9780080505640}.
\newblock URL
  \url{https://www.taylorfrancis.com/books/mono/10.4324/9780080505640/energy-simulation-building-design-joseph-clarke}.

\bibitem[Klein et~al.(2007)Klein, Beckman, Mitchell, Duffie, Duffie, Freeman,
  Mitchell, Braun, Evans, Kummer, et~al.]{klein2007trnsys}
SA~Klein, WA~Beckman, JW~Mitchell, JA~Duffie, NA~Duffie, TL~Freeman,
  JC~Mitchell, JE~Braun, B~Evans, J~Kummer, et~al.
\newblock Trnsys 16: A transient system simulation program: mathematical
  reference.
\newblock \emph{Trnsys}, 5:\penalty0 389--396, 2007.

\bibitem[Crawley et~al.(2000)Crawley, Lawrie, Pedersen, and
  Winkelmann]{crawley2000energy}
Drury~B Crawley, Linda~K Lawrie, Curtis~O Pedersen, and Frederick~C Winkelmann.
\newblock Energy plus: energy simulation program.
\newblock \emph{ASHRAE journal}, 42\penalty0 (4):\penalty0 49--56, 2000.

\bibitem[Sonta et~al.(2018)Sonta, Simmons, and Jain]{Sonta.2018}
Andrew~J. Sonta, Perry~E. Simmons, and Rishee~K. Jain.
\newblock Understanding building occupant activities at scale: An integrated
  knowledge-based and data-driven approach.
\newblock \emph{Advanced Engineering Informatics}, 37:\penalty0 1--13, 2018.
\newblock ISSN 14740346.
\newblock \doi{10.1016/j.aei.2018.04.009}.

\bibitem[Sousa(2012)]{Sousa.2012}
Joana Sousa.
\newblock Energy simulation software for buildings review and comparison.
\newblock In \emph{International Workshop on Information Technology for Energy
  Applicatons-IT4Energy, Lisabon}, 2012.

\bibitem[Nguyen et~al.(2014)Nguyen, Reiter, and Rigo]{Nguyen.2014}
Anh-Tuan Nguyen, Sigrid Reiter, and Philippe Rigo.
\newblock A review on simulation-based optimization methods applied to building
  performance analysis.
\newblock \emph{Applied Energy}, 113:\penalty0 1043--1058, 2014.
\newblock ISSN 03062619.
\newblock \doi{10.1016/j.apenergy.2013.08.061}.

\bibitem[Seyedzadeh et~al.(2018)Seyedzadeh, Rahimian, Glesk, and
  Roper]{Seyedzadeh.2018}
Saleh Seyedzadeh, Farzad~Pour Rahimian, Ivan Glesk, and Marc Roper.
\newblock Machine learning for estimation of building energy consumption and
  performance: a review.
\newblock \emph{Visualization in Engineering}, 6\penalty0 (1), 2018.
\newblock \doi{10.1186/s40327-018-0064-7}.

\bibitem[Westermann and Evins(2019)]{Westermann.2019}
Paul Westermann and Ralph Evins.
\newblock Surrogate modelling for sustainable building design -- a review.
\newblock \emph{Energy and Buildings}, 198:\penalty0 170--186, 2019.
\newblock ISSN 03787788.
\newblock \doi{10.1016/j.enbuild.2019.05.057}.

\bibitem[Chakraborty and Elzarka(2019)]{Chakraborty.2019}
Debaditya Chakraborty and Hazem Elzarka.
\newblock Advanced machine learning techniques for building performance
  simulation: a comparative analysis.
\newblock \emph{Journal of Building Performance Simulation}, 12\penalty0
  (2):\penalty0 193--207, 2019.
\newblock ISSN 1940-1493.

\bibitem[Deb et~al.(2017)Deb, Zhang, Yang, Lee, and Shah]{Deb.2017}
Chirag Deb, Fan Zhang, Junjing Yang, Siew~Eang Lee, and Kwok~Wei Shah.
\newblock A review on time series forecasting techniques for building energy
  consumption.
\newblock \emph{Renewable and Sustainable Energy Reviews}, 74:\penalty0
  902--924, 2017.
\newblock ISSN 13640321.

\bibitem[Banihashemi et~al.(2017)Banihashemi, Ding, and Wang]{Banihashemi.2017}
Saeed Banihashemi, Grace Ding, and Jack Wang.
\newblock Developing a hybrid model of prediction and classification algorithms
  for building energy consumption.
\newblock \emph{Energy Procedia}, 110:\penalty0 371--376, 2017.
\newblock ISSN 18766102.
\newblock \doi{10.1016/j.egypro.2017.03.155}.

\bibitem[Amasyali and El-Gohary(2018)]{Amasyali.2018}
Kadir Amasyali and Nora~M. El-Gohary.
\newblock A review of data-driven building energy consumption prediction
  studies.
\newblock \emph{Renewable and Sustainable Energy Reviews}, 81:\penalty0
  1192--1205, 2018.
\newblock ISSN 13640321.
\newblock \doi{10.1016/j.rser.2017.04.095}.

\bibitem[Coakley et~al.(2011)Coakley, Raftery, Molloy, and White]{Coakley.2011}
Daniel Coakley, Paul Raftery, Padraig Molloy, and Gearoid White.
\newblock Calibration of a detailed bes model to measured data using an
  evidence-based analytical optimisation approach.
\newblock 2011.

\bibitem[Hsu(2015)]{Hsu.2015}
David Hsu.
\newblock Identifying key variables and interactions in statistical models of
  building energy consumption using regularization.
\newblock \emph{Energy}, 83:\penalty0 144--155, 2015.
\newblock ISSN 03605442.
\newblock \doi{10.1016/j.energy.2015.02.008}.

\bibitem[Borrmann et~al.(2018)Borrmann, K{\"o}nig, Koch, and
  Beetz]{Borrmann.2018}
Andr{\'e} Borrmann, Markus K{\"o}nig, Christian Koch, and Jakob Beetz.
\newblock Building information modeling: Why? what? how?
\newblock In \emph{Building information modeling}, pages 1--24. Springer, 2018.

\bibitem[Farzaneh et~al.(2019)Farzaneh, Monfet, and Forgues]{Farzaneh.2019}
Aida Farzaneh, Danielle Monfet, and Daniel Forgues.
\newblock Review of using building information modeling for building energy
  modeling during the design process.
\newblock \emph{Journal of Building Engineering}, 23:\penalty0 127--135, 2019.
\newblock ISSN 23527102.
\newblock \doi{10.1016/j.jobe.2019.01.029}.

\bibitem[Abualdenien and Borrmann(2018)]{Abualdenien.2018}
J.~Abualdenien and A.~Borrmann.
\newblock Multi-lod model for describing uncertainty and checking requirements
  in different design stages.
\newblock In Jan Karlsh{\o}j and Raimar Scherer, editors, \emph{eWork and
  eBusiness in Architecture, Engineering and Construction}, pages 187--195.
  {CRC Press}, 2018.
\newblock ISBN 9780429506215.
\newblock \doi{10.1201/9780429506215-24}.

\bibitem[Latiffi et~al.(2015)Latiffi, Brahim, Mohd, and Fathi]{Latiffi.2015}
Aryani~Ahmad Latiffi, Juliana Brahim, Suzila Mohd, and Mohamad~Syazli Fathi.
\newblock Building information modeling (bim): Exploring level of development
  (lod) in construction projects.
\newblock \emph{Applied Mechanics and Materials}, 773-774:\penalty0 933--937,
  2015.
\newblock \doi{10.4028/www.scientific.net/AMM.773-774.933}.

\bibitem[Singh and Geyer(2020)]{Singh.2020}
Manav~Mahan Singh and Philipp Geyer.
\newblock Information requirements for multi-level-of-development bim using
  sensitivity analysis for energy performance.
\newblock \emph{Advanced Engineering Informatics}, 43:\penalty0 101026, 2020.
\newblock ISSN 14740346.

\bibitem[Gao et~al.(2019)Gao, Koch, and Wu]{Gao.2019}
Hao Gao, Christian Koch, and Yupeng Wu.
\newblock Building information modelling based building energy modelling: A
  review.
\newblock \emph{Applied Energy}, 238:\penalty0 320--343, 2019.
\newblock ISSN 03062619.
\newblock \doi{10.1016/j.apenergy.2019.01.032}.

\bibitem[Nutkiewicz et~al.(2018)Nutkiewicz, Yang, and Jain]{Nutkiewicz.2018}
Alex Nutkiewicz, Zheng Yang, and Rishee~K. Jain.
\newblock Data-driven urban energy simulation (due-s): A framework for
  integrating engineering simulation and machine learning methods in a
  multi-scale urban energy modeling workflow.
\newblock \emph{Applied Energy}, 225:\penalty0 1176--1189, 2018.
\newblock ISSN 03062619.
\newblock \doi{10.1016/j.apenergy.2018.05.023}.

\bibitem[Tian et~al.(2018)Tian, Heo, de~Wilde, Li, {Da Yan}, Park, Feng, and
  Augenbroe]{Tian.2018}
Wei Tian, Yeonsook Heo, Pieter de~Wilde, Zhanyong Li, {Da Yan}, Cheol~Soo Park,
  Xiaohang Feng, and Godfried Augenbroe.
\newblock A review of uncertainty analysis in building energy assessment.
\newblock \emph{Renewable and Sustainable Energy Reviews}, 93:\penalty0
  285--301, 2018.
\newblock ISSN 13640321.
\newblock \doi{10.1016/j.rser.2018.05.029}.

\bibitem[Hyndman and Athanasopoulos(2018)]{Hyndman.2018}
Rob~J. Hyndman and George Athanasopoulos.
\newblock Forecasting: Principles and practice.
\newblock 2018.

\bibitem[Cleveland et~al.(1990)Cleveland, Cleveland, McRae, and
  Terpenning]{Cleveland.1990}
Robert~B. Cleveland, William~S. Cleveland, Jean~E. McRae, and Irma Terpenning.
\newblock Stl: A seasonal-trend decomposition.
\newblock \emph{Journal of official statistics}, 6\penalty0 (1):\penalty0
  3--73, 1990.

\bibitem[{Der Kiureghian} and Ditlevsen(2009)]{DerKiureghian.2009}
Armen {Der Kiureghian} and Ove Ditlevsen.
\newblock Aleatory or epistemic? does it matter?
\newblock \emph{Structural safety}, 31\penalty0 (2):\penalty0 105--112, 2009.

\bibitem[{Jeremiah Liu} et~al.(2019){Jeremiah Liu}, {John Paisley},
  {Marianthi-Anna Kioumourtzoglou}, and {Brent Coull}]{JeremiahLiu.2019}
{Jeremiah Liu}, {John Paisley}, {Marianthi-Anna Kioumourtzoglou}, and {Brent
  Coull}.
\newblock Accurate uncertainty estimation and decomposition in ensemble
  learning.
\newblock 2019.

\bibitem[Mili{\'c} et~al.(2018)Mili{\'c}, Ekel{\"o}w, and Moshfegh]{Milic.2018}
Vlatko Mili{\'c}, Klas Ekel{\"o}w, and Bahram Moshfegh.
\newblock On the performance of lcc optimization software opera-milp by
  comparison with building energy simulation software ida ice.
\newblock \emph{Building and Environment}, 128:\penalty0 305--319, 2018.
\newblock ISSN 03601323.
\newblock \doi{10.1016/j.buildenv.2017.11.012}.

\bibitem[Judea(2021)]{Judea.2021}
Pearl Judea.
\newblock Radical empiricism and machine learning research.
\newblock \emph{Journal of Causal Inference}, 9\penalty0 (1):\penalty0 78--82,
  2021.
\newblock URL
  \url{https://EconPapers.repec.org/RePEc:bpj:causin:v:9:y:2021:i:1:p:78-82:n:2}.

\bibitem[Monteiro et~al.(2018)Monteiro, Costa, Pina, Santos, and
  Ferr{\~a}o]{Monteiro.2018}
Claudia~Sousa Monteiro, Carlos Costa, Andr{\'e} Pina, Maribel~Y. Santos, and
  Paulo Ferr{\~a}o.
\newblock An urban building database (ubd) supporting a smart city information
  system.
\newblock \emph{Energy and Buildings}, 158:\penalty0 244--260, 2018.
\newblock ISSN 03787788.
\newblock \doi{10.1016/j.enbuild.2017.10.009}.

\bibitem[Mantha et~al.(2016)Mantha, Menassa, and Kamat]{Mantha.2016}
Bharadwaj~R.K. Mantha, Carol~C. Menassa, and Vineet~R. Kamat.
\newblock A taxonomy of data types and data collection methods for building
  energy monitoring and performance simulation.
\newblock \emph{Advances in Building Energy Research}, 10\penalty0
  (2):\penalty0 263--293, 2016.
\newblock ISSN 1751-2549.
\newblock \doi{10.1080/17512549.2015.1103665}.

\bibitem[de~Jaeger et~al.(2020)de~Jaeger, Reynders, Callebaut, and
  Saelens]{Jaeger.2020}
Ina de~Jaeger, Glenn Reynders, Chadija Callebaut, and Dirk Saelens.
\newblock A building clustering approach for urban energy simulations.
\newblock \emph{Energy and Buildings}, 208:\penalty0 109671, 2020.
\newblock ISSN 03787788.

\bibitem[Bilal et~al.(2016)Bilal, Oyedele, Qadir, Munir, Ajayi, Akinade,
  Owolabi, Alaka, and Pasha]{Bilal.2016}
Muhammad Bilal, Lukumon~O. Oyedele, Junaid Qadir, Kamran Munir, Saheed~O.
  Ajayi, Olugbenga~O. Akinade, Hakeem~A. Owolabi, Hafiz~A. Alaka, and Maruf
  Pasha.
\newblock Big data in the construction industry: A review of present status,
  opportunities, and future trends.
\newblock \emph{Advanced Engineering Informatics}, 30\penalty0 (3):\penalty0
  500--521, 2016.
\newblock ISSN 14740346.
\newblock \doi{10.1016/j.aei.2016.07.001}.

\bibitem[{\O}sterg{\aa}rd et~al.(2016){\O}sterg{\aa}rd, Jensen, and
  Maagaard]{stergard.2016}
Torben {\O}sterg{\aa}rd, Rasmus~L. Jensen, and Steffen~E. Maagaard.
\newblock Building simulations supporting decision making in early design -- a
  review.
\newblock \emph{Renewable and Sustainable Energy Reviews}, 61:\penalty0
  187--201, 2016.
\newblock ISSN 13640321.
\newblock \doi{10.1016/j.rser.2016.03.045}.

\bibitem[Wang and Augenbroe(2017)]{Wang.2017}
Qinpeng Wang and Godfried Augenbroe.
\newblock Combined sensitivity ranking of input parameters and model forms of
  building energy simulation.
\newblock In \emph{Proceedings of BS2017: 15th Conference of International
  Building Performance Simulation Association}, pages 3--9, 2017.

\bibitem[Kamel and Memari(2019)]{Kamel.2019}
Ehsan Kamel and Ali~M. Memari.
\newblock Review of bim's application in energy simulation: Tools, issues, and
  solutions.
\newblock \emph{Automation in Construction}, 97:\penalty0 164--180, 2019.
\newblock ISSN 09265805.
\newblock \doi{10.1016/j.autcon.2018.11.008}.

\bibitem[Roman et~al.(2020)Roman, Bre, Fachinotti, and Lamberts]{Roman.2020}
Nadia~D. Roman, Facundo Bre, Victor~D. Fachinotti, and Roberto Lamberts.
\newblock Application and characterization of metamodels based on artificial
  neural networks for building performance simulation: A systematic review.
\newblock \emph{Energy and Buildings}, 217:\penalty0 109972, 2020.
\newblock ISSN 03787788.
\newblock \doi{10.1016/j.enbuild.2020.109972}.

\bibitem[Roman()]{Roman.2020b}
Nadia Roman.
\newblock Data for: Application and characterization of metamodels based on
  artificial neural networks for building performance simulation: a systematic
  review.

\bibitem[Papadopoulos et~al.(2018)Papadopoulos, Azar, Woon, and
  Kontokosta]{Papadopoulos.2018}
Sokratis Papadopoulos, Elie Azar, Wei-Lee Woon, and Constantine~E. Kontokosta.
\newblock Evaluation of tree-based ensemble learning algorithms for building
  energy performance estimation.
\newblock \emph{Journal of Building Performance Simulation}, 11\penalty0
  (3):\penalty0 322--332, 2018.
\newblock ISSN 1940-1493.
\newblock \doi{10.1080/19401493.2017.1354919}.

\bibitem[Marsland(2015)]{Marsland.2015}
Stephen Marsland.
\newblock \emph{Machine learning: an algorithmic perspective}.
\newblock {CRC Press}, 2015.

\bibitem[Arjunan et~al.(2020)Arjunan, Poolla, and Miller]{Arjunan.2020}
Pandarasamy Arjunan, Kameshwar Poolla, and Clayton Miller.
\newblock Energystar++: Towards more accurate and explanatory building energy
  benchmarking.
\newblock \emph{Applied Energy}, 276:\penalty0 115413, 2020.
\newblock ISSN 03062619.
\newblock \doi{10.1016/j.apenergy.2020.115413}.
\newblock URL \url{http://arxiv.org/pdf/1910.14563v2}.

\bibitem[Polikar(2006)]{Polikar.2006}
R.~Polikar.
\newblock Ensemble based systems in decision making: Ieee circuits and systems
  magazine, 6(3), 21-45.
\newblock \emph{IEEE Circuits and Systems Magazine}, 6\penalty0 (3):\penalty0
  21--45, 2006.
\newblock ISSN 1531-636X.
\newblock \doi{10.1109/MCAS.2006.1688199}.

\bibitem[{Guolin Ke} et~al.(2017){Guolin Ke}, {Qi Meng}, {Thomas Finley},
  {Taifeng Wang}, {Wei Chen}, {Weidong Ma}, {Qiwei Ye}, and {Tie-Yan
  Liu}]{GuolinKe.2017}
{Guolin Ke}, {Qi Meng}, {Thomas Finley}, {Taifeng Wang}, {Wei Chen}, {Weidong
  Ma}, {Qiwei Ye}, and {Tie-Yan Liu}.
\newblock Lightgbm: A highly efficient gradient boosting decision tree.
\newblock 2017.

\bibitem[Chen and Guestrin()]{Chen.201639}
Tianqi Chen and Carlos Guestrin.
\newblock Xgboost: A scalable tree boosting system.
\newblock URL \url{https://arxiv.org/pdf/1603.02754}.

\bibitem[{Tong Xiao} et~al.(2022){Tong Xiao}, {Peng Xu}, {Ruikai He}, and
  {Huajing Sha}]{TongXiao.2022}
{Tong Xiao}, {Peng Xu}, {Ruikai He}, and {Huajing Sha}.
\newblock Status quo and opportunities for building energy prediction in
  limited data context---overview from a competition.
\newblock \emph{Applied Energy}, 305:\penalty0 117829, 2022.
\newblock ISSN 03062619.
\newblock \doi{10.1016/j.apenergy.2021.117829}.
\newblock URL
  \url{https://www.sciencedirect.com/science/article/pii/S0306261921011570}.

\bibitem[Vogt et~al.(2018)Vogt, Remmen, Lauster, Fuchs, and
  M{\"u}ller]{Vogt.2018}
Marcus Vogt, Peter Remmen, Moritz Lauster, Marcus Fuchs, and Dirk M{\"u}ller.
\newblock Selecting statistical indices for calibrating building energy models.
\newblock \emph{Building and Environment}, 144:\penalty0 94--107, 2018.
\newblock ISSN 03601323.
\newblock \doi{10.1016/j.buildenv.2018.07.052}.

\end{thebibliography}






\end{document}